\documentclass[12pt]{article}
\usepackage{amsfonts}
\usepackage{amssymb}
\usepackage{amsthm}
\usepackage{hyperref}
\usepackage{amssymb,amsmath,amsthm}
\usepackage{enumerate}

\newcommand{\sss}{\setcounter{equation}{0}}
\newtheorem{theorem}{THEOREM}[section]

\newtheorem{lemma}[theorem]{LEMMA}

\newtheorem{corollary}[theorem]{COROLLARY}

\newcommand{\ere}{ {\mathbb R}}

\def\beq{\begin{equation}}
\def\ene{\end{equation}}

\def \ds {\displaystyle}
\newcommand{\bull}{\hfill $\Box$}

\def\qed{\ifhmode\unskip\nobreak\fi\ifmmode\ifinner
\else\hskip5pt\fi\fi\hbox{\hskip5pt\vrule width4pt height6pt
depth1.5pt\hskip1pt}}

\def\var{\varepsilon}

\def\e{\mathbf E}
\def \b{\mathbf B}
\def\d{\mathbf D}
\def\h{\mathbf H}
\def\x{\mathbf x}
\def\y{\mathbf y}
\def\c{\mathbf c}
\def\H{\mathcal H}
\def\C{\mathbf C}
\def\X{\underline{\mathbf x}}
\def\Y{\underline{\mathbf y}}
\begin{document}
\baselineskip=20 pt
\parskip 6 pt

\title{A Rigorous Time-Domain  Analysis  of Full--Wave
Electromagnetic  Cloaking (Invisibility)
\thanks{ PACS classification scheme 2006: 41.20.Jb, 02.30.Tb,02.30.Zz, 02.60.Lj.} \thanks{ Research partially
supported by  CONACYT under Project P42553­F.}}
 \author{  Ricardo Weder\thanks{ †Fellow Sistema Nacional de Investigadores.}
\\Instituto de Investigaciones en Matem\'aticas Aplicadas y en Sistemas \\
Departamento de M\'etodos Matem\'aticos y Num\'ericos\\Universidad Nacional Aut\'onoma de
M\'exico \\
Apartado Postal 20-726, M\'exico DF 01000
\\ weder@servidor.unam.mx}

\date{}
\maketitle
\begin{center}
\begin{minipage}{5.75in}
\centerline{{\bf Abstract}}
\bigskip
There is currently a great deal of interest in the theoretical and practical possibility of cloaking objects
from the observation by electromagnetic
waves. The basic idea of these invisibility devices \cite{glu1, glu2, le}, \cite{pss1}  is to use
anisotropic
{\it transformation media} whose permittivity and permeability $\var^{\lambda\nu},  \mu^{\lambda\nu}$,   are
obtained from the ones,
$\var_0^{\lambda\nu}, \mu^{\lambda\nu}_0$, of   isotropic media,  by  singular transformations of coordinates.

In this paper  we study electromagnetic cloaking in the time-domain using the formalism of time-dependent scattering theory \cite{rsIII}.
  This formalism provides us with a rigorous method to analyze   the propagation of electromagnetic wave packets with finite energy in
  {\it transformation media}. In particular, it allows us to settle in an unambiguous  way the mathematical
   problems posed by the singularities of the
  inverse of the permittivity and the permeability of the {\it transformation media} on the boundary of the
   cloaked objects.  Von Neumann's theory of
  self-adjoint
  extensions of symmetric operators plays an important role on this issue.
We write Maxwell's equations in Schr\"odinger form with the
electromagnetic propagator playing the role of the Hamiltonian. We
prove that the electromagnetic propagator outside of the cloaked
objects is essentially self-adjoint. This means that it has only one
self-adjoint extension, $A_\Omega$, and that this self-adjoint
extension generates the only possible unitary time evolution, with
constant energy, for finite energy electromagnetic waves, propagating outside of
the cloaked objects.
\end{minipage}
\begin{minipage}{5.75in}

Moreover, $A_\Omega$ is unitarily equivalent
to the electromagnetic propagator in the medium
$\var_0^{\lambda\nu}, \mu^{\lambda\nu}_0$. Using this fact, and
since the coordinate transformation is the identity outside of a
ball, we prove that the scattering operator is the identity. This
implies that for any incoming finite-energy electromagnetic wave
packet the outgoing wave packet is precisely the same.
In other words, it is not possible to detect the cloaked objects in
any scattering experiment where a finite-energy wave packet is sent
towards the cloaked objects, since the outgoing wave packet that is
measured after  interaction  is the same as the incoming one. Our
results give a rigorous proof that the construction of  \cite{glu1,
glu2, le}, \cite{pss1}  cloaks passive and active devices
from observation by electromagnetic waves. Actually, the cloaking
outside is independent of what is inside  the cloaked objects.

As is well known, self-adjoint extensions can be understood in terms
of boundary conditions. Actually, for the electromagnetic fields in
the domain of $A_\Omega$ the component tangential to the exterior of the boundary of
the cloaked objects of both, the electric and the magnetic field
have to be zero. This boundary condition is self-adjoint in our case
because  the  permittivity and the permeability  are degenerate on the
boundary of the cloaked objects.

Furthermore, we prove cloaking for general anisotropic materials. In
particular, our results prove that it is possible to cloak objects
inside general crystals.

\end{minipage}
\end{center}


\section{Introduction}\sss
There is currently a great deal of interest in the theoretical and practical possibility of cloaking objects from the observation by electromagnetic
fields. The basic idea of these invisibility devices \cite{glu1, glu2, le}, \cite{pss1}  is to use anisotropic
{\it transformation media} whose permittivity and permeability, $\var^{\lambda\nu},  \mu^{\lambda\nu}$,  are obtained from the ones,
$\var_0^{\lambda\nu}, \mu^{\lambda\nu}_0$,
of   isotropic media,  by  singular transformations of coordinates. The singularities lie on the boundary of the objects to be cloaked. Here
the {\it material interpretation} is taken. Namely, the   $\var^{\lambda\nu}, \mu^{\lambda\nu}$ and the $\var_0^{\lambda\nu},  \mu^{\lambda\nu}_0$,
represent the components in flat Cartesian space of the permittivity and the permeability  of physical media  with {\it different material properties}.
It appears that with existing technology it is possible to construct media as described above using artificially structured metamaterials. In
\cite{glu1, glu2}  a proof of cloaking was given for the conductivity equation -i.e., in the case of zero frequency- from detection by measurement of the
 Dirichlet to Neumann map that relates the value of the electric potential on the boundary to its normal derivative. The papers  \cite{le} and \cite{pss1}
 consider electromagnetic waves in the geometrical optics approximation, i.e. for large frequencies. In
 \cite{smjcpss} a experimental verification of
 cloaking is presented and \cite{ccks} and \cite{cpss} give a numerical simulation. A rigorous prof of cloaking
 has already been given by  \cite{gklu} where
 fixed frequency waves were studied, i.e., in the frequency domain. They consider a class of finite energy solutions
 to Maxwell's equations in a bounded set, $O$, that  contains the cloaked  object on its interior, and they
 prove cloaking, at any frequency,  with
 respect to the measurement of the Cauchy data of these solutions    on the boundary of $O$. We give further
 comments on this paper  below.
For other results on this problem see \cite{sps} and \cite{lp}. In   \cite{mbw}
  cloaking of elastic waves is considered, and the history of invisibility is discussed.

In this paper we study electromagnetic cloaking in the time-domain using the formalism of time-dependent scattering theory \cite{rsIII}.
This formalism provides us with a rigorous method to analyze   the propagation of electromagnetic wave packets with finite energy in
{\it transformation media}. In particular, it allows us to settle in an unambiguous way the mathematical problems posed by the singularities of the
inverse of the permittivity and the permeability of the {\it transformation media} on the boundary of the cloaked objects. Von Neumann's theory of
self-adjoint extensions of symmetric operators plays an important role on this issue.
We write Maxwell's equations in Schr\"odinger form with the electromagnetic propagator playing the role of the
Hamiltonian. We prove that the electromagnetic propagator outside of the cloaked
objects is essentially self-adjoint. This means that it has only one
self-adjoint extension, $A_\Omega$, and that this self-adjoint
extension generates the only possible unitary time evolution, with
constant energy, for finite energy electromagnetic waves propagating outside of
the cloaked objects.
Moreover, $A_\Omega$ is unitarily equivalent
to the electromagnetic propagator in the medium
$\var_0^{\lambda\nu}, \mu^{\lambda\nu}_0$. Using this fact, and
since the coordinate transformation is the identity outside of a
ball, we prove that the scattering operator is the identity. This
implies that for any incoming finite-energy electromagnetic wave
packet the outgoing wave packet is precisely the same.
In other words, it is not possible to detect the cloaked objects in
any scattering experiment where a finite-energy wave packet is sent
towards the cloaked objects, since the outgoing wave packet that is
measured after  interaction  is the same as the incoming one. Our
results give a rigorous proof that the construction of  \cite{glu1,
glu2, le}, \cite{pss1}  cloaks passive and active devices
from observation by electromagnetic waves. Actually, the cloaking
outside is independent of what is inside  the cloaked objects.

As is well known, self-adjoint extensions can be understood in terms
of boundary conditions. Actually, for the electromagnetic fields in
the domain of $A_\Omega$ the component tangential to the exterior of the boundary of
the cloaked objects of both, the electric and the magnetic field
have to be zero. This boundary condition is self-adjoint in our case
because  the  permittivity and the permeability  are degenerate on the
boundary of the cloaked objects.

Furthermore, we prove cloaking for general anisotropic materials. In
particular, our results prove that it is possible to cloak objects
inside general crystals.

Even though, as mentioned above, the cloaking  is independent of the  cloaked objects, and in particular,
the cloaking outside is not affected by the presence of passive and/or active devices inside the cloaked objects,
we discuss the dynamics of electromagnetic waves inside the cloaked objects for completeness, since it helps to
understand  the above mentioned independence of cloaking from the properties of the cloaked objects.

We prove that every self-adjoint extension of the electromagnetic propagator in a {\it transformation medium}
is the direct sum of the  unique self-adjoint extension in the exterior of the cloaked objects, $A_\Omega$,
with some self-adjoint extension of the electromagnetic propagator in the interior of the cloaked objects.
Each of these self-adjoint extensions corresponds to a possible unitary time evolution for finite energy
electromagnetic waves. As is well known, the fact that time evolution is unitary assures us that energy is
conserved. This results implies  that the electromagnetic waves inside and outside of the cloaked  objects completely
decouple from each other. Actually, the  electromagnetic waves inside  the cloaked objects are not allowed to
leave them, and viceversa, the electromagnetic waves
outside can not go inside.

In terms of boundary conditions, this means that transmission conditions that link the electromagnetic fields
inside and outside the cloaked objects are not allowed, since they do not correspond to self-adjoint extensions
of the electromagnetic propagator, and then, they do not lead to a unitary dynamics that conserves  energy.
Furthermore, choosing a particular self-adjoint extension of the electromagnetic propagator of
the cloaked objects  amounts to choosing
some boundary condition on the inside of the boundary of the cloaked objects.
In other words, any possible unitary dynamics implies the existence of some boundary condition   on the inside
of the boundary of the cloaked objects.

The fact that there is a large class of self-adjoint extensions -or boundary conditions- that can be taken inside
the cloaked objects could be useful in order to enhance cloaking in practice, where one has to consider
{\it approximate transformation media } as well as in the analysis of the stability of cloaking.

Actually, we consider a slightly more general construction than the one of  \cite{glu1, glu2,le}, \cite{pss1}
since we allow for a finite number of cloaked objects.

In \cite{gklu} a very general construction for cloaking is introduced. In the case of Maxwell's equations all
their constructions are made within the context of the permittivity and the permeability tensor densities being conformal to each other, i.e.,
 multiples
of each other by a positive scalar function. In particular, all isotropic media are included in this category.
They mention that both for mathematical
and practical reasons  it would be very interesting to understand cloaking for general anisotropic materials in the absence of this assumption.
In this paper we actually solve this problem, since we prove cloaking for all general anisotropic materials.
In particular, our results prove that it is possible to cloak objects inside general crystals.

Note, moreover, that  \cite{gklu} also considers the cases of the  Helmholtz equation. We do not
discuss this problems here.

Furthermore, remark that the existing theorems in the uniqueness of inverse scattering do not apply under the present conditions.


In  \cite{gklu}  cloaking is proven with respect to  the Cauchy data at any fixed
frequency  given on a surface that encloses the cloaked object. In the case where
the permittivity and the permeability are bounded  above and below it is well known that the Cauchy  data at a
fixed frequency is equivalent to  the scattering matrix at the same frequency. See for example \cite{n}
and \cite{u}. This equivalence is, however, not proven in the case where the permittivity and the permeability
are degenerate on the boundary of the objects. In fact, it is perhaps even not true for general degenerate media
that are not transformation media since in this case it is possible that there are finite energy
electromagnetic waves that  are absorbed by the boundary of the objects as
$ t\rightarrow \pm \infty$. If this is true, the equivalence will not hold since the Cauchy data in a surface
that encloses the objects will not contain information on the waves that are asymptotically absorbed by the
boundary of the objects. It is a   problem of independent interest to see if
this actually happens or not for general degenerate permittivities and permeabilities. For an example of
scattering by a  bounded obstacle with a singular boundary and Neumann
boundary condition, where this happens see \cite{hw}. For a similar situation in the scattering of
electromagnetic waves by a Schwarzschild black-hole see \cite{ba}.
Note that in  our approach we directly consider  the scattering operator that is measured in scattering
 experiments.

In the analysis of Maxwell's equations with permittivity and permeability that are independent of frequency the dispersion of the medium is not
taken into account. This means that cloaking will hold for electromagnetic wave packets with a narrow enough
range of frequencies, such that this
assumption is valid.

The paper is organized as follows. In Section 2 we prove our results in electromagnetic cloaking. In Section 3
we consider the propagation of electromagnetic waves in the interior of the cloaked objects. In Section 4
we formulate cloaking as a boundary value problem outside of the cloaked objects for the Maxwell equations
at a fixed frequency, following our analysis of the self-adjoint extensions of the electromagnetic propagator.
In particular, we give the appropriate boundary condition on the outside of the boundary of the cloaked objects.
Finally, in Section
5 we prove  cloaking of infinite cylinders. This is of interest since this is the case considered in the
experimental verification in \cite{smjcpss} and in the numerical simulations of  \cite{ccks} and \cite{cpss}.
Of course, \cite{smjcpss} only consider a slice of the cylinder. In Sections 3 and 4  we give further
comments on the results of \cite{gklu}.

\noindent {\bf Addendum}

\noindent After the previous version of this paper was posted in the
arXiv we published  the  paper \cite{we3} where we generalized the
results of this paper on spherical cloaks to the case of high-order
cloaks, and where we also discussed cloaking in the frequency
domain. Moreover, in our paper \cite{we4} we identified the {\it
cloaking boundary condition } that has to be satisfied in the inside
of the boundary of the cloaked objects, in the case where the
permittivity and the permeability are bounded above and below inside
the cloaked objects.

\section{Electromagnetic Cloaking}
\sss
Let us consider Maxwell's equations,

\begin{eqnarray}
\nabla \times \e &=& -\frac{\partial}{\partial t}\b, \,\, \nabla \times \h \,=\,\frac{\partial }{\partial t} \d,
\label{1.1} \\\nonumber\\
 \nabla \cdot \b&=&0, \nabla \cdot \d\,=
 \,0,
\label{1.2}
 \end{eqnarray}

 in a domain, $\Omega  \subset \mathbf \ere^3$, as follows,

\beq \Omega:= \mathbf \ere^3 \setminus \cup_{j=1}^N K_j, \,\,
K_j\cap K_l =\emptyset, j \neq l \label{1.3} \ene where $K_j,
j=1,2,\cdots,N,$ are the objects to be cloaked. We assume that each
$K_j$ is a ball  with center ${\mathbf c}_j$ and radius $a_j$, i.e.,

\beq
K_j=\left\{ \x \in \ere^3: |\x- {\mathbf c}_j| \leq a_j\right\}, j=1,2,\cdots,N.
\label{1.4}
\ene

The cloaked objects are denoted by
$$
K:= \cup_{j=1}^N K_j.
$$

We  designate the Cartesian coordinates of  $\x$  by  $ x^\lambda, \lambda=1,2,3$ and by $E_\lambda, H_\lambda, B^\lambda, D^\lambda, \lambda=1,2,3$,
 respectively, the components of $ \e,\h,\b$, and $\d$. As usual, we denote by $\varepsilon^{\lambda\nu}$ and $\mu^{\lambda\nu}$, respectively,
the permittivity and the permeability. We have that,

\beq
D^\lambda= \varepsilon^{\lambda\nu}E_\nu, \,\,\, B^\lambda= \mu^{\lambda \nu}H_\nu,
\label{1.6}
\ene
where we use the standard convention of summing over repeated lower and upper indices.

We consider now a transformation from $\Omega_0:=\ere^3 \setminus
\{\c_1, \c_2,\cdots,\c_N\}$ onto $\Omega$ that was first used   to obtain cloaking for the
conductivity equation, i.e. at zero frequency, by \cite{glu1,glu2} and then by
\cite{pss1} for cloaking  electromagnetic waves (for a
related result in two dimensions using conformal mappings see
\cite{le}).



For any $\y \in \ere^3$ we denote, $\hat{\y}:= \y /|\y|$.  Let
$y^\lambda, \lambda=1,2,3,$ designate the cartesian coordinates of $ \y \in \Omega_0$.
Take $ b_j > a_j, j=1,2,\cdots,N$. Then, for $0 < |\y-\c_j| \leq b_j$,  we define,

\beq \x=\x(\y) = f(\y):= \c_j + \left( \frac{b_j-a_j}{b_j}|\y-\c_j|+a_j\right) \,\, \widehat{\y-\c_j}.
\label{1.8}
\ene
Note that this transformation blows
up the point $\c_j$ onto $\partial K_j$ and that it sends the
punctuated ball $\tilde{B}_{\ds \c_j}(b_j):= \left\{ \y \in
\ere^3: 0 <|\y-\c_j|\leq b_j\right\}$ onto  the spherical shell,  $ a_j< |\x-\mathbf c_j| \leq  b_j  $. We
assume that,

\beq \tilde{B}_{\ds\c_j}(b_j) \cap
\tilde{B}_{\ds\c_l}(b_l)=\emptyset,\, j\neq l, 1 \leq j,l \leq N.
\label{1.9} \ene For $\y \in \ere^3 \setminus \cup_{j=1}^N
\tilde{B}_{\ds \c_j}(b_j)$ we define the transformation to be
the identity, $\x=\x(\y)=f(\y):= \y$. Our transformation is a
bijection from $\Omega_0$ onto $\Omega$.  By $ \y=\y(\x):=
f^{-1}(\x)$ we designate the inverse transformation. We denote the
elements of the Jacobian matrix by $ A^\lambda_{\lambda'}$, \beq
A^\lambda_{\lambda'}:= \frac{\partial x^\lambda}{\partial
y^{\lambda'}}. \label{1.10} \ene
 Note that the
$A^\lambda_{\lambda'}\in C^1\left(\Omega_0 \setminus \cup_{j=1}^N \partial \tilde{B}_{\ds \c_j}(b_j)\right)$. We designate by $A^{\lambda'}_\lambda$ the
elements
of the Jacobian of the inverse bijection,
$\y=\y(\x):=f^{-1}(\x)$,

\beq
A^{\lambda'}_{\lambda}:= \frac{\partial y^{\lambda'}}{\partial x^{\lambda}}\in C^1\left(\Omega
\setminus \cup_{j=1}^N \partial \tilde{B}_{\ds \c_j}(b_j)\right).
\label{1.11}
\ene
The papers  \cite{glu1,glu2} and  \cite{pss1} considered the  case where $N=1,
\c_1=0$.

We take here the so called {\it material interpretation} and we consider our transformation as a bijection between two different
spaces, $\Omega_0$ and $\Omega$. However, our transformation can be considered, as well, as a change of coordinates in $\Omega_0$. Of
course, these two point of view are mathematically equivalent. This means, in particular, that under our transformation the Maxwell
equations in $\Omega_0$ and in $\Omega$ will have the same invariance  that they have under change of coordinates in
three-space. See, for example, \cite{po}. Let us denote by $\Delta$ the determinant of the Jacobian matrix (\ref{1.10}). Then,
\beq
\Delta= \frac{b_j-a_j}{b_j}\left( \frac{\frac{b_j-a_j}{b_j}|\y-\c_j|+a_j}{|\y-\c_j|}\right)^2, \, \hbox{for}\, 0< |\y-\c_j| \leq b_j.
\label{1.12}
\ene
This result is easily obtained rotating into a coordinate system such that, $\y-\c_j=( |\y-\c_j|,0,0)$
\cite{sps}. For $\y \in \Omega_0 \setminus
\cup_{j=1}^N  \tilde{B}_{\ds \c_j}(b_j), \Delta \equiv 1$.

Let us denote by $\e_0,\h_0,\b_0, \d_0,  \varepsilon^{\lambda\nu}_0,
\mu^{\lambda\nu}_0$, respectively, the electric and magnetic fields, the magnetic induction, the electric displacement, and the permittivity
and permeability of $\Omega_0$. $ \varepsilon^{\lambda\nu}_0,
\mu^{\lambda\nu}_0$, are positive, Hermitian matrices that are constant in $\Omega_0$.

The electric field is a covariant vector that transforms as,

\beq
E_\lambda(\x) = A_\lambda^{\lambda'}(\y)E_{0,\lambda'}(\y).
\label{1.13}
\ene

The magnetic field $\h$ is a covariant pseudo-vector, but as we only consider space transformations with positive determinant, it also transforms
as in (\ref{1.13}).
The magnetic induction $\b$ and the electric displacement $\d$ are contravariant vector densities of weight one that transform as

\beq
B^\lambda(\x) =  \left(\Delta (\y)\right)^{-1}  A_{\lambda'}^{\lambda}(\y) B^{\lambda'}_0(\y),
\label{1.14}
\ene
with the same transformation for $\d$.
The permittivity and permeability are contravariant tensor densities of weight one that transform as,
\beq
\varepsilon^{\lambda\nu}(\x)=  \left(\Delta (\y)\right)^{-1} A^{\lambda}_{\lambda'}(\y)\, A^{\nu}_{\nu'}(\y)\, \varepsilon^{\lambda' \nu'}_0(\y),
\label{1.15}
\ene
with the same transformation for $ \mu^{\lambda\nu}$. The Maxwell equations (\ref{1.1}, \ref{1.2}) are the same in both spaces $\Omega$ and $\Omega_0$.
Let us denote by $\varepsilon_{\lambda \nu}, \mu_{\lambda \nu}, \varepsilon_{0\lambda \nu}, \mu_{0\lambda \nu}$, respectively, the inverses of
the corresponding permittivity and permeability. They are covariant tensor densities of weight minus one that transform as,

\beq
\varepsilon_{\lambda\nu}(\x)=  \Delta (\y) A^{\lambda'}_{\lambda}(\y)\, A^{\nu'}_{\nu}(\y)\, \varepsilon_{0\lambda' \nu'}(\y),\,
\mu_{\lambda\nu}(\x)=  \Delta (\y) A^{\lambda'}_{\lambda}(\y)\, A^{\nu'}_{\nu}(\y)\, \mu_{0\lambda' \nu'}(\y).
\label{1.16}
\ene
 Note that
\beq
 \det \varepsilon^{\lambda\nu}=\Delta^{-1} \det \varepsilon^{\lambda\nu}_0,\, \det \mu^{\lambda\nu}=\Delta^{-1} \det \mu^{\lambda\nu}_0,\,
\label{1.17}
 \ene

\beq
\det \varepsilon_{\lambda\nu}=\Delta \det \varepsilon_{0\lambda\nu},\,  \det \mu_{\lambda\nu}=\Delta \det \mu_{0\lambda\nu}.
\label{1.18}
\ene

We now introduce the Hilbert spaces of electric and magnetic fields with  finite energy. The  $\e_0,\h_0,\b_0, \d_0$, were defined in $\Omega_0$,
but since $\ere^3\setminus \Omega_0 =\{\c_j\}_{j=1}^N$ is of measure zero, we can consider them as defined in $\ere^3$, what we do below.

We denote by $\H_{0E}$ the Hilbert space of all
measurable, square integrable, $\C^3-$ \,valued functions defined on $\ere^3$ with the scalar product,
\beq
\left(\e^{(1)}_0, \e^{(2)}_0\right)_{0 E}:= \int_{\ere^3}E^{(1)}_{0\lambda} \, \var^{\lambda\nu}_0\,
\overline{ E^{(2)}_{0\nu}}\, d\y^3.
\label{1.19}
\ene
We similarly define the Hilbert space,$\H_{0H}$,  of all
measurable, square integrable, \, $\C^3-$ valued functions defined on $\ere^3$ with the scalar product,
\beq
\left(\h^{(1)}_0, \h^{(2)}_0\right)_{0 H}:=  \int_{\ere^3}H^{(1)}_{0\lambda} \, \mu^{\lambda\nu}_0\,\overline{H^{(2)}_{0\nu}}\, d\y^3.
\label{1.20}
\ene

The Hilbert space of finite energy fields in $\ere^3$ is the direct sum

\beq
\H_0:= \H_{0 E}\oplus \H_{0 H}.
\label{1.21}
\ene

 Moreover, we designate  by $\H_{\Omega E}$ the Hilbert space of all
measurable, $\C^3-$ valued functions defined on $\Omega$  that are square integrable with the weight $\var^{\lambda\nu}$, with the scalar product,
\beq
\left(\e^{(1)}, \e^{(2)}\right)_{\Omega E}:= \int_{\Omega}E^{(1)}_\lambda\,\var^{\lambda\nu}\,\overline{E^{(2)}_\nu}\, d\x^3.
\label{1.22}
\ene

 Finally, we denote  by $\H_{\Omega H}$ the Hilbert space of all
measurable, $\C^3-$ valued functions defined on $\Omega$  that are square integrable with the weight $\mu^{\lambda\nu}$, with the scalar product,
\beq
\left(\h^{(1)}, \h^{(2)}\right)_{\Omega H}:= \int_{\Omega}H^{(1)}_\lambda\,\mu^{\lambda\nu}\,\overline{H^{(2)}_\nu}\, d\x^3.
\label{1.23}
\ene

The Hilbert space of finite energy fields in $\Omega$ is the direct sum

\beq
\H_\Omega:= \H_{\Omega E}\oplus \H_{\Omega H}.
\label{1.24}
\ene

We now write the Maxwell's equations (\ref{1.1}) in  Schr\"odinger form. We first consider the case of $\ere^3$. We denote by
$ \var_0$ and $\mathbf \mu_0$, respectively, the matrices with entries $\var_{0\lambda\nu}$ and $\mu_{0\lambda\nu}$. Recall that
$\left(\nabla\times \e\right)^\lambda=
s^{\lambda\nu\rho}\frac{\partial}{\partial x_\nu }E_\rho $ where $s^{\lambda\nu\rho}$ is the
permutation
contravariant pseudo-density of weight $-1$ (see section 6 of chapter II of \cite{po}, where a different notation is used). By $a_0$ we denote
 the
following formal differential operator,

\beq
a_0 \left(\begin{array}{c}\e_0\\ \h_0\end{array}\right)=i \left(\begin{array}{c}
\var_0 \nabla\times \h_0\\- \mu_0 \nabla\times \e_0\end{array}\right).
\label{1.24b}
\ene
Here, as usual, we denote, $\var_0 \nabla\times \h_0:= \var_{0\lambda\nu} (\nabla \times \h_0)^\nu$, and   $\mu_0 \nabla\times \e_0=\mu_{0\lambda\nu}
(\nabla \times \e_0)^\nu$.
Then, equations (\ref{1.1}) are equivalent to,
\beq
i\frac{\partial}{\partial t}\left(\begin{array}{c}\e_0\\\h_0\end{array}\right)= a_0\left(\begin{array}{c}\e_0\\\h_0\end{array}\right).
\label{1.25}
\ene

Let us denote by $\C^1_0(\ere^3)$ the set of all $\C^6-$valued continuously differentiable functions on $\ere^3$ that have compact support.
Then, $a_0$ with domain
$\C^1_0(\ere^3)$ is a symmetric operator in $\H_0$, i.e., $ a_0 \subset a_0^\ast$. Moreover, it is essentially self-adjoint in $\H_0$, i.e.,
it has only one self-adjoint extension, that we denote by $A_0$. Its domain is given by,

\beq
D(A_0)=\left\{ \left(\begin{array}{c}\e_0\\\h_0\end{array}\right) : a_0 \left(\begin{array}{c}\e_0\\\h_0\end{array}\right) \in \H_0\right\},
\label{1.26}
\ene
and,
\beq
A_0 \left(\begin{array}{c}\e_0 \\\h_0\end{array}\right)= a_0 \left(\begin{array}{c}\e_0\\\h_0\end{array}\right),\,
\left(\begin{array}{c}\e_0\\\h_0\end{array}\right)
\in D(A_0),
\label{1.27}
\ene
where the derivatives are taken in distribution sense.
These results follow easily from the fact that -via the Fourier transform-  $a_0$ is unitarily equivalent to multiplication by a
matrix valued function  that is symmetric with respect to the scalar product of $\H_0$. Moreover, it follows from explicit computation that
the only eigenvalue of $A_0$ is zero, that it has infinite multiplicity, and that,

\beq
\H_{0\perp}:= \left(\hbox{kernel}\, A_0\right)^\perp = \left\{\left(\begin{array}{c}\e_0\\\h_0\end{array}\right)\in \H_0:
\frac{\partial}{\partial x_\lambda}
\var^{\lambda\nu}_0E_{0\nu}=0 ,  \frac{\partial}{\partial x_\lambda}
\mu^{\lambda\nu}_0H_{0\nu}=0\right\}.
\label{1.28}
\ene
Furthermore, $A_0$ has no singular-continuous spectrum and its absolutely-continuous spectrum is $\ere$. See, for example, \cite{we1,we2}.

Taking any
\beq
\left(\begin{array}{c}\e_0 \\\h_0\end{array}\right)\in \H_{0\perp}\cap D(A_0)
\label{1.29}
\ene
we obtain a finite energy solution to the Maxwell equations (\ref{1.1}, \ref{1.2}) as follows

\beq
    \left(\begin{array}{c}\e_0 \\\h_0\end{array}\right)(t) =e^{-it A_0} \left(\begin{array}{c}\e_0 \\\h_0\end{array}\right).
\label{1.29b}
\ene
This is the unique finite energy solution with initial value at $t=0$ given by (\ref{1.29}). Note that as $e^{-itA_0} \H_{0\perp}\subset \H_{0\perp}$
equations (\ref{1.2}) are satisfied for all times if they are satisfied at $t=0$.

Let us now consider the case of $\Omega$.  We denote by
$ \var$ and $\mathbf \mu$, respectively, the matrices with entries $\var_{\lambda\nu}$ and $\mu_{\lambda\nu}$.

We now define the following formal differential operator,
\beq
a_\Omega \left(\begin{array}{c}\e\\ \h\end{array}\right)=i \left(\begin{array}{c}
\var \nabla\times \h\\- \mu \nabla\times \e\end{array}\right).
\label{1.30}
\ene

Equations (\ref{1.1}) are equivalent to,
$$
i\frac{\partial}{\partial t}\left(\begin{array}{c}\e\\\h\end{array}\right)= a_\Omega\left(\begin{array}{c}\e\\\h\end{array}\right).
$$

Let us denote by $\C^1_0(\Omega)$ the set of all ${\C}^6-$valued continuously differentiable functions on $\Omega$ that have compact support.
Then, $a_\Omega$ with domain $\C^1_0(\Omega)$ is a symmetric operator in $\H_\Omega$. To construct a unitary dynamics that preserves energy we
have to analyse
the self-adjoint extensions of $a_\Omega$.

We denote by $U_E$ the following unitary operator from $\H_{0 E}$ onto $\H_{\Omega E}$,

\beq
\left(U_E \e_0\right)_\lambda(\x): = A^{\lambda'}_\lambda  E_{0\lambda'}(\y),
\label{1.31}
\ene
and by $U_H$ the unitary operator from $\H_{0H}$ onto $\H_{\Omega H}$,

\beq
\left(U_H \h_0\right)_\lambda(\x): = A^{\lambda'}_\lambda  H_{0\lambda'}(\y).
\label{1.32}
\ene

Then,
\beq
U:= U_E \oplus U_H
\label{1.33}
\ene
is a unitary operator from $\H_0$ onto $\H_\Omega$.


We denote by $a_{00}$ the restriction of $a_0$ to $\C^1_0(\Omega_0)$. The operator $a_{00}$ is essentially self-adjoint and its only self-adjoint
extension is $A_0$. This follows from the essential self-adjointness of $a_0$ and from the fact that any function in $\C^1_0(\ere^3)$ can be
approximated in the graph norm of $a_0$ by functions in $\C^1_0(\Omega_0)$.
To prove this take any continuously differentiable real-valued
function, $\phi$,
defined on $\ere$ such that, $\phi(y)=0, |y|\leq 1$ and $\phi(y)=1, |y| \geq 2$. Then, for any

$$
\left(\begin{array}{c}\e_0\\\h_0\end{array}\right) \in \C^1_0(\ere^3),
$$
we have that,
$$
     \prod_{j=1}^ N \phi (n|\y-\c_j|)\, \left(\begin{array}{c}\e_0\\\h_0\end{array}\right) \in \C^1_0(\Omega_0)
$$
and moreover,
$$
\hbox{s-}\lim_{n\rightarrow \infty}  \prod_{j=1}^ N \phi (n|\y-\c_j|)\, \left(\begin{array}{c}\e_0\\\h_0\end{array}\right)=  \left(\begin{array}{c}\e_0\\\h_0\end{array}\right),
$$
$$
\hbox{s-}\lim_{n\rightarrow \infty} a_0  \prod_{j=1}^ N \phi (n|\y-\c_j|)  \, \left(\begin{array}{c}\e_0\\\h_0\end{array}\right)=
a_0 \left(\begin{array}{c}\e_0\\\h_0\end{array}\right),
$$
where by $\hbox{s-}\lim$ we designate the strong limit in $\H_0$.

As $a_{00}$ is essentially self-adjoint, it follows from the invariance of Maxwell equations that $a_\Omega$ is essentially self-adjoint, and that its
unique self-adjoint extension, that we denote by $A_\Omega$, satisfies

\beq
A_\Omega = U\, A_0\, U^\ast.
\label{1.35}
\ene
For the proof of these facts see \cite{we3}.
Hence, we have the following theorem.
\begin{theorem}
The operator $a_\Omega$ is essentially self-adjoint, and its unique self-adjoint extension, $A_\Omega$, satisfies (\ref{1.35}).
\end{theorem}
The unitary equivalence given by (\ref{1.35}) implies that $A_\Omega$ has the same spectral properties that $A_0$. Namely, it has no
singular-continuous spectrum, the absolutely-continuous spectrum is $\ere$ and the only eigenvalue is zero and it has infinite multiplicity.
Moreover,

\beq
\H_{\Omega \perp}:= \left(\hbox{kernel}\, A_\Omega\right)^\perp = \left\{\left(\begin{array}{c}\e\\\h\end{array}\right)\in \H_\Omega:
\frac{\partial}{\partial x_\lambda}
\var^{\lambda\nu}E_{\nu}=0 ,  \frac{\partial}{\partial x_\lambda}
\mu^{\lambda\nu}H_{\nu}=0\right\}.
\label{1.36}
\ene
Furthermore, taking any
\beq
\left(\begin{array}{c}\e \\\h\end{array}\right)\in \H_{\Omega\perp}\cap D(A_\Omega)
\label{1.37}
\ene
we obtain a finite energy solution to the Maxwell equations (\ref{1.1}, \ref{1.2}) as follows

\beq
    \left(\begin{array}{c}\e \\\h\end{array}\right)(t) =e^{-it A_\Omega} \left(\begin{array}{c}\e \\\h\end{array}\right).
\label{1.37b}
\ene
This is the unique finite energy solution with initial value at $t=0$ given by (\ref{1.37}). Note that as $e^{-itA_\Omega} \H_{\Omega\perp}\subset
\H_{\Omega\perp}$ equations (\ref{1.2}) are satisfied for all times if they are satisfied at $t=0$. We can consider more general solutions by considering the scale of spaces associated with $A_\Omega$, but we do not go into this
direction here.

The facts that $a_\Omega$ is essentially self-adjoint and that its unique self-adjoint extension $A_\Omega$ is unitarily equivalent to the
propagator $A_0$ of the homogeneous medium are
strong statements. They mean that the only possible unitary dynamics in $\Omega$ that preserves energy is given by (\ref{1.37b}) and that this dynamics
 is unitarily equivalent
to the free dynamics in $\ere^3$ given by (\ref{1.29b}). In fact, $\partial \Omega$ acts like a horizon for electromagnetic waves propagating in
$\Omega$ in the sense that the dynamics is uniquely defined without any need to consider  the cloaked objects $K=\cup_{j=1}^N K_j$.
As we will prove below this implies electromagnetic cloaking for all frequencies in the strong sense that the scattering operator
is the identity.

Since $  D(A_\Omega)=U D(A_0) $, for any $(\e,\h)^T \in D(A_\Omega)$ there is a $(\e_0,\h_0)^T \in
D(A_0)$ such that

\beq
\left(\begin{array}{c}\e\\ \h\end{array}\right)= U \left(\begin{array}{c}\e_0\\ \h_0\end{array}\right).
\label{2.1}
\ene
Then, it follows from (\ref{1.31}, \ref{1.32}, \ref{1.33}) that

\beq
\e\times \mathbf n=0, \h\times  \mathbf n=0, \,\, \hbox {in }\,\, \partial K_+,
\label{2.2}
\ene
where ${\partial K}_+$ denotes the outside of the boundary of the cloaked objects, $K$, and $\mathbf n$ is the
normal vector to $\partial K_+$, if $(\e_0,\h_0)$ are, for example, bounded near $\partial K_+$. That is to say, for electromagnetic fields in the domain of $A_\Omega$ the tangential
components of both, the electric and the magnetic field vanish in the exterior of the boundary of the cloaked objects. This is a self-adjoint boundary condition
 because the permittivity   and the permeability are degenerate on $\partial K_+$.

Let $\chi_\Omega$ be the characteristic function of $\Omega$, i.e., $\chi_\Omega(\x)=1
, \x \in \Omega, \chi_\Omega(\x)=0, \x \in \ere^3\setminus \Omega$.
We define,

\beq
\left(J \left(\begin{array}{c}\e_0\\ \h_0\end{array}\right)\right)(\x):=  \chi_\Omega(\x) \left(\begin{array}{c}\e_0\\ \h_0\end{array}\right)(\x).
\label{1.55}
\ene

By (\ref{1.8}, \ref{1.12}, \ref{1.15}),
$$
\left| \var^{\lambda\nu}(\x)\right| \leq C, \quad  \left| \mu^{\lambda\nu}(\x)\right| \leq C, \quad \x\in \Omega.
$$
Then, $J$ is a bounded operator from $\H_0$ into $\H_\Omega$.

The wave operators are defined as follows,
\beq
W_\pm = \hbox{s-}\lim
_{t \rightarrow \pm \infty} e^{itA_\Omega}\, J e^{-itA_0} P_{0\perp},
\label{1.56}
\ene
where $P_{0\perp}$ denotes the projector onto $\H_{0\perp}$.

Let us designate by $\mathbf W^{1,2}(\ere^3)$ the Sobolev space of $\mathbf C^6$ valued functions.
We denote by $I$ the identity operator on $\H_0$.
Then,
\begin{lemma}
\beq
W_{\pm}= U P_{0\perp}.
\label{1.57}
\ene
\end{lemma}
\noindent {\it Proof:} Denote,

$$
W(t):= e^{itA_\Omega}\, J \, e^{-itA_0} P_{0\perp}.
$$
By (\ref{1.35}), for any $\varphi \in \H_0$
\beq
W(t)\varphi = \psi(t)+ U P_{0\perp}\varphi,
\label{1.58}
\ene
with
\beq
\psi(t):= U \,e^{itA_0}\, \left(U^\ast J-I \right)\, e^{-itA_0} P_{0\perp}\varphi.
\label{2.3}
\ene

Let $B_R$ denote the ball of center zero and radius $R$ in $\ere^3$. Since for $|y|\geq R$, with $R$ large enough, our transformation, $ \x=f(\y)$,
 is the identity,
$\x=\y$, and in consequence, $A^\lambda_{\lambda'}(\y)= \delta^{\lambda}_{\lambda'}$ for $|\y| \geq R$, we have that,

\beq
\left(U^\ast J -I\right)= \left(U^\ast J -I\right) \chi_{\ds B_R}.
\label{2.4}
\ene
It follows that,

\beq
\hbox{s-}\lim_{t \rightarrow \pm \infty} \psi(t)= U\, \hbox{s-}\lim_{t \rightarrow \pm \infty} e^{itA_0}\vartheta(t)
\label{2.5}
\ene
with,
\beq
\vartheta(t):=  \left(U^\ast J -I\right) \chi_{\ds B_R} e^{-itA_0} P_{0\perp} \varphi.
\label{2.6}
\ene

We have that,

\beq
\left\|\vartheta(t)\right\|_{\H_0}  \leq  \left\| J \chi_{\ds B_R} e^{-itA_0} P_{0\perp}\varphi\right\|_{\H}
+ \left\|\chi_{\ds B_R} e^{-itA_0} P_{0\perp}\varphi\right\|_{\H_0} \leq C  \left\|\chi_{\ds B_R} e^{-itA_0} P_{0\perp}\varphi\right\|_{\H_0}.
\label{1.59}
\ene

Then, as  $(A_0+i)^{-1} P_{0\perp}$
is bounded from $\H_0$ into $\mathbf W^{1,2}(\ere^3)$  \cite{we1} \cite{we2}, it follows from the Rellich local compactness theorem that

$$
\chi_{\ds B_R} \, (A_0+i)^{-1} P_{0\perp}
$$
is a compact operator in $\H_0$.  Suppose that $ \varphi \in D(A_0) \cap \H_{0\perp}$. Then,

\beq
\hbox{s-} \lim_{t \rightarrow \pm \infty} \chi_{\ds B_R} e^{-itA_0} P_{0\perp}\varphi =
\hbox{s-} \lim_{t \rightarrow \pm \infty} \chi_{\ds B_R}   (A_0+i)^{-1} P_{0\perp}  e^{-itA_0} (A_0+i) \varphi=0,
\label{2.7}
\ene
and whence, by (\ref{1.59}),
\beq
\hbox{s-}\lim_{t \rightarrow \pm \infty} \vartheta(t)= 0,
\label{2.8}
\ene
and it follows that in this case,

\beq
  \hbox{s-}\lim_{t \rightarrow \pm \infty}    \psi(t)=0.
\label{1.60}
\ene
By continuity this is also true for $\varphi \in \H_{0\perp}$.

Then, (\ref{1.57}) follows from (\ref{1.58}) and (\ref{1.60}).

\bull

The scattering operator is defined as

\beq
S:= W_+^\ast\, W_-.
\label{2.9}
\ene
\begin{corollary}
\beq
S=P_{0\perp}.
\label{1.61}
\ene
\end{corollary}
\noindent {\it Proof:} This is immediate from (\ref{1.57}) because $U^\ast\, U=I$.

\bull

Let us denote by $S_\perp$ the restriction of $S$ to $\H_{0\perp}$. $S_{\perp}$ is the physically relevant scattering operator that acts in the
Hilbert space $\H_{0\perp}$ of finite energy fields that satisfy equations (\ref{1.2}). We designate by $I_\perp$ the identity operator on
$\H_{0\perp}$. We have that,

\begin{corollary}
\beq
S_\perp= I_\perp.
\label{1.62}
\ene
\end{corollary}
\noindent {\it Proof:} This follows from Corollary 2.3.

\bull

The fact that $S_\perp$ is the identity operator on $\H_{0\perp}$  means that there is perfect cloaking for
all frequencies. Suppose that for
 very
negative times we are given an incoming wave packet
$e^{-it A_0}\varphi_-$, with
$\varphi_- \in \H_{0\perp}$. Then, for large positive times the outgoing wave packet is given by $e^{-it A_0}\varphi_+$ with
$\varphi_+= S_\perp\varphi_-$.
But, as $S_\perp=I$, we have that $\varphi_+= \varphi_-$ and then,

$$
e^{-it A_0}\varphi_- = e^{-it A_0}\varphi_+.
$$

Since the incoming and the outgoing wave packets are the same there is no way to detect the cloaked objects $K$ from scattering experiments
performed in
$\Omega$.

In this paper we considered {\it transformation media} obtained from a singular transformation that blows up a
 finite number of points, by simplicity,
and since this is the situation in the applications. Suppose that we have a transformation that is singular
in a set of points that we  call
$M$ and denote now $\Omega_0:= \ere^3 \setminus M$. What we really used in the proofs is that
$\mathbf W^{1,2}(\ere^3)= \mathbf W^{1,2}_0(\Omega_0)$ where $\mathbf W^{1,2}_{0}(\Omega_0)$ denotes the
completion of $\mathbf C^\infty_0(\Omega_0)$ in the norm of $\mathbf W^{1,2}(\ere^3)$. We also assumed
that $\var_0^{\lambda\nu}, \mu^{\lambda\nu}_0$ are constant. What was actually  needed  is that $a_0$ is
essentially self-adjoint. All our results
hold under this more general conditions provided that in (\ref{1.56}, \ref{1.57}) and (\ref{1.61}) we replace $P_{0\perp}$
by the projector onto the absolutely-continuous subspace of $A_0$ and that we assume that
$D(A_0) \cap \H_{0ac} \subset \mathbf W^{1,2}(\ere^3)$, where we
have denoted the absolutely-continuous
subspace of $A_0$ by  $ \H_{0ac}$. Moreover, $S_\perp$ has to be defined as the restriction of
$S$ to $\H_{0ac}$ and in (\ref{1.62}) $I_\perp$ has to
be the identity operator on $\H_{0ac}$. Note that under these general assumptions $A_0$ could have non-zero eigenvalues and singular-continuous
spectrum.

For example, $\mathbf W^{1,2}(\ere^3)= \mathbf W^{1,2}_0(\Omega_0)$ if $M$   has zero Sobolev one capacity \cite{af,kkm, km}.
Moreover, assume that the permittivity and the permeability tensor densities $\var_0^{\lambda\nu},
 \mu^{\lambda\nu}_0$ are
bounded below and above. Under this condition   $a_0$ is  essentially self-adjoint.
Furthermore, let us denote by $\hat{\H}_0$ the Hilbert space of finite energy solutions defined as in
(\ref{1.21}) but with $\varepsilon_0^{\lambda\nu}= \mu_0^{\lambda\mu}= \delta^{\lambda\mu} $. Let $ \hat{A}_0,
\hat{\H}_{0\perp}$ be, respectively, the electromagnetic propagator in $\hat{\H}_{0}$ and the orthogonal
complement of its kernel. We have that $\H_{0}$ and $\hat{\H}_{0}$ are the same set of functions with equivalent norms.
Furthermore, $ D(A_0)= D(\hat{A}_0),\hbox{kernel}\, \hat{A}_0= \hbox{kernel}\, A_0$. Moreover, $(\e_0,\h_0)^T\in
\H_{0\perp}$
if and only if $ \e_0= \varepsilon_0 \hat{\e}_0, \h_0= \mu_0 \hat{\h}_0$ for some $(\hat{\e}_0,\hat{\h}_0)\in
\hat{\H}_{0\perp}$.
As \cite{we1,we2} $D(\hat{A}_0)\cap \hat{\H}_{0\perp} \subset \mathbf W^{1,2}(\ere^3)$ we have
 that $D(A_0) \cap \H_{0\perp} \subset \mathbf W^{1,2}(\ere^3)$
if $ \varepsilon_0, \mu_0$ are bounded operators on $\mathbf W^{1,2}(\ere^3)$ and this is true if the derivatives
$\frac{\partial}{\partial y_\rho}\varepsilon_{0}, \frac{\partial}{\partial y_\rho}\mu_{0}$
are bounded operators on $\hat{\H}_0$ for $\rho =1,2,3$. Note, furthermore, that $\H_{0ac} \subset \H_{0\perp}$.

\section{Electromagnetic Waves Inside the Cloaked Objects }
\sss
Let us now consider  the propagation of electromagnetic  waves in the cloaked objects.
For this purpose we assume that in each $K_j$ the permittivity and the permeability are given by $\var^{\lambda\nu}_j, \mu^{\lambda\nu}_j$, with
inverses
$\var_{j\lambda\nu}, \mu_{j\lambda\nu}$ and where $\var_j, \mu_j$ are the matrices with entries $\var_{j\lambda\nu}, \mu_{j\lambda\nu}$.
Furthermore, we assume that $  0 < \var^{\lambda\nu}_j, \mu^{\lambda \nu}_j \leq C, \x \in K_j$ and that for any compact set $Q$ contained in the interior
of $K_j$ there is a positive constant $C_Q$ such that $\det \var^{\lambda\nu}_j > C_Q, \det\mu^{\lambda \nu}_j > C_Q, \x \in Q$. In other words, we only
allow for possible
singularities of $\var_j, \mu_j$ on the boundary of $K_j$.

We designate  by $\H_{j E}$ the Hilbert space of all
measurable, $\C^3-$ valued functions defined on $K_j$  that are square integrable with the weight $\var^{\lambda\nu}_j$, with the scalar product,
\beq
\left(\e^{(1)}_j, \e^{(2)}_j\right)_{j E}:= \int_{K_j}E^{(1)}_{j\lambda}\,\var^{\lambda\nu}_j\,\overline{E^{(2)}_{j\nu}}\, d\x^3.
\label{1.38}
\ene

 Similarly, we denote  by $\H_{j H}$ the Hilbert space of all
measurable, $\C^3-$ valued functions defined on $K_j$  that are square integrable with the weight $\mu^{\lambda\nu}_j$, with the scalar product,
\beq
\left(\h^{(1)}_j, \h^{(2)}_j\right)_{j H}:= \int_{K_j}H^{(1)}_{j\lambda}\,\mu^{\lambda\nu}_j\,\overline{H^{(2)}_{j\nu}}\, d\x^3.
\label{1.39}
\ene

The Hilbert space of finite energy fields in $K_j$ is the direct sum

\beq
\H_{j}:= \H_{j E}\oplus \H_{j H},
\label{1.40}
\ene
and the Hilbert space in the cloaked objects $K$ is the direct sum,

$$
\H_K:=  \oplus_{j=1}^N \H_{j}.
$$

The complete Hilbert space of finite energy fields  including the cloaked objects is,

\beq
\H:= \H_\Omega\oplus \H_K.
\label{1.41}
\ene

We now write (\ref{1.1}) as a Schr\"odinger equation in each $K_j$ as before.
We define the following formal differential operator,
\beq
a_j \left(\begin{array}{c}\e_j\\ \h_j\end{array}\right)=i \left(\begin{array}{c}
\var_j \nabla\times \h_j\\- \mu_j \nabla\times \e_j\end{array}\right).
\label{1.42}
\ene

Equation (\ref{1.1}) in $K_j$ is equivalent to
\beq
i\frac{\partial}{\partial t}\left(\begin{array}{c}\e_j\\\h_j\end{array}\right)= a_j\left(\begin{array}{c}\e_j\\\h_j\end{array}\right).
\label{1.43}
\ene
Let us denote by $\C^1_0(\hat{K}_j)$ the set of all ${\C}^6-$valued continuously differentiable functions on $K_j$ that have compact support in the
interior of
$K_j$, that we denote by $\hat{K}_j:= K_j \setminus \partial K_j$.
 Then, $a_j$ with domain $C^1_0(\hat{K}_j)$ is a symmetric operator in $\H_{j}$.
We denote,

\beq
a:= a_\Omega \oplus_{j=1}^N a_j,
\label{1.44}
\ene
with domain,
\beq
D(a):= \left\{ \left(\begin{array}{c}\e_\Omega\\ \h_\Omega\end{array}\right)\oplus_{j=1}^N \left(\begin{array}{c}\e_j\\ \h_j\end{array}\right)
\in \C^1_0(\Omega)\oplus_{j=1}^N \C^1_0(\hat{K}_j)\right\}.
\label{1.45}
\ene
The operator $a$ is symmetric in $\H$. The possible unitary dynamics that preserve energy for the whole system including the cloaked objects
$K$ are given by the self-adjoint extensions of $a$. Let us denote $\overline{a}$ the closure of $a$, with similar notation for $a_\Omega,
a_j, j=1,\cdots,N$. Then,
$$
\overline{a}= A_\Omega \oplus_{j=1}^N \overline{a_j},
$$
where we used the fact that as $a_\Omega$ is essentially self-adjoint, $\overline{a_\Omega}=A_\Omega
$.
The adjoint of $a$ is given by,

\beq
D(a^\ast)= \left\{ \left(\begin{array}{c}\e_\Omega\\ \h_\Omega\end{array}\right)\oplus_{j=1}^N \left(\begin{array}{c}\e_j\\ \h_j\end{array}\right)
\in \H :    \left(\begin{array}{c}\e_\Omega\\ \h_\Omega\end{array}\right) \in D(A_\Omega), a_j \left(\begin{array}{c}\e_j\\ \h_j\end{array}\right)
\in \H_j \right\},
\label{1.46}
\ene
and
\beq
a^\ast \left( \left(\begin{array}{c}\e_\Omega\\ \h_\Omega\end{array}\right)\oplus_{j=1}^N \left(\begin{array}{c}\e_j\\ \h_j\end{array}\right)\right)
= A_\Omega \left(\begin{array}{c}\e_\Omega\\ \h_\Omega\end{array}\right) \oplus_{j=1}^N a_j \left(\begin{array}{c}\e_j\\ \h_j\end{array}\right),
\label{1.47}
\ene
for
\beq
\left(\begin{array}{c}\e_\Omega\\ \h_\Omega\end{array}\right)\oplus_{j=1}^N \left(\begin{array}{c}\e_j\\ \h_j\end{array} \right)\in
D(a^\ast).
\label{1.48}
\ene

Let us denote by ${\mathcal K}_{\Omega \pm}:= \hbox{kernel}(i  \mp a_\Omega^\ast), {\mathcal K}_{j \pm}:= \hbox{kernel}(i  \mp a_j^\ast)$
the deficiency subspaces of $ a_\Omega$ and $a_j, j=1,\cdots,N$. Since $a_\Omega$ is  essentially self-adjoint
$\mathcal K_{\Omega \pm} =\{0\}$. Let  $\mathcal K_\pm:= \oplus_{j=1}^N \mathcal K_{j\pm}$ be the deficiency subspaces of
$ a_K:=\oplus_{j=1}^N a_j$. Suppose that $\mathcal K_\pm$ have the same dimension. Then, it  follows from Corollary 1 in page 141 of \cite{rsII}
that there is a one-to-one correspondence between   self-adjoint extensions
of $a_K$ and unitary maps
from $\mathcal K_+$  into $\mathcal K_-$. If $V$ is such a unitary, then the corresponding  self-adjoint extension $A_{KV}$ is given by,
$$
D(A_{K V})=\left\{ \varphi +\varphi_+ +V\varphi_+ : \varphi \in D(\overline{a_K}), \varphi_+ \in \mathcal K_+\right\},
$$
and
$$
A_K \varphi = \overline{a_K}\varphi+ i\varphi_+ -i V\varphi_+.
$$
Hence, since $\mathcal K_{\Omega \pm} =\{0\}$ and $\overline{a}= A_\Omega \oplus \overline{a_K}$ there is a one-to-one correspondence  between
self-adjoint extensions
of $a$ and unitary maps, $V$, from $\mathcal K_+$  into $\mathcal K_-$. The self-adjoint extension  $A_V$  corresponding to $V$ is given by,
$$
A_V =A_\Omega\oplus A_{KV}.
$$

Thus, we have proven the following theorem.

\begin{theorem}
Every self-adjoint extension, $A$, of $a$ is the direct sum of  $A_\Omega$ and of some self-adjoint extension, $A_K$ of $a_K$, i.e.,
\beq
A= A_\Omega \oplus A_K.
\label{1.49}
\ene
\end{theorem}
This theorem tells us that the cloaked objects $K$ and the exterior $\Omega$ are completely decoupled and that
  we are free to choose any boundary
condition inside the cloaked objects $K$ that makes $a_K$ self-adjoint without disturbing the cloaking
effect in $\Omega$. Boundary conditions that make
$A_K$ self-adjoint are well known. See for example, \cite{pi1}, \cite{pi2}, \cite{lei} and \cite{bs}.

It follows from explicit computation that zero is an eigenvalue of every $A_K$ with infinite multiplicity and that,

\beq
\H_{K \perp}:= \left(\hbox{kernel}\, A_K\right)^\perp \subset \left\{\left(\begin{array}{c}\e\\\h\end{array}\right)\in \H_K:
\frac{\partial}{\partial x_\lambda}
\var^{\lambda\nu}_K E_{\nu}=0 ,  \frac{\partial}{\partial x_\lambda}
\mu^{\lambda\nu}_K H_{\nu}=0\right\},
\label{1.50}
\ene
where by $ \var^{\lambda\nu}_K(\x):= \var^{\lambda\nu}_j(\x)$ for $\x \in K_j$, and
$ \mu^{\lambda\nu}_K(\x):= \mu^{\lambda\nu}_j(\x)$ for $\x \in K_j, j=1,2,\cdots,N$.
It follows that zero is an eigenvalue of $A$ with infinite multiplicity and that,
\beq
\H_{ \perp}:= \left(\hbox{kernel}\, A\right)^\perp=  \H_{\Omega \perp}\oplus \H_{K \perp}.
\label{1.51}
\ene
For any $\varphi= \varphi_\Omega\oplus\varphi_K \in \H_{ \perp} \cap D(A)$,

\beq
e^{-itA}\varphi= e^{-it A_\Omega}\, \varphi_\Omega \oplus e^{-itA_K}\,\varphi_K
\label{1.52}
\ene
is the unique solution of Maxwell's equations (\ref{1.1}, \ref{1.2}) with finite energy that is equal to $\varphi$ at $t=0$. This shows once again
that the
dynamics in $\Omega$ and in $K$ are completely decoupled. If at $t=0$ the electromagnetic fields are zero in $\Omega$, they remain equal to zero
for all
times, and viceversa. Actually, electromagnetic waves inside  the cloaked objects are not allowed to leave them, and viceversa, electromagnetic waves
outside can not go inside. This implies, in particular, that the presence of active devices inside the cloaked objects  has no effect on the cloaking
outside.
In terms of boundary conditions, this means that transmission conditions that link the electromagnetic fields
inside and outside the cloaked objects are not allowed.
Furthermore, choosing a particular self-adjoint extension of the electromagnetic propagator of
the cloaked objects  amounts to choosing
some boundary condition on the inside of the boundary of the cloaked objects.
In other words, any possible unitary dynamics implies the existence of some boundary condition   on the inside
of the boundary of the cloaked objects.
The particular boundary condition that nature  will take  depends on the specific properties of the
metamaterial used to build the transformation media as well us on the properties of the media inside the
cloaked objects. Note that this does not mean that we have to put any physical surface, a lining, on the
surface of the cloaked object to enforce any particular boundary condition  on the inside, since as we already
mentioned this plays no role in  the cloaking outside. It would be, however, of theoretical interest to see
what the interior boundary condition turns out to be for specific cloaked objects and metamaterials.
These results apply to the {\it exact transformation media} that we consider on this paper.
However, the fact that there is a large class of self-adjoint extensions -or boundary conditions- that can be
taken inside the cloaked objects could be useful in order to enhance cloaking in practice, where one has to consider
{\it approximate transformation media } as well as in the analysis of the stability of cloaking.

The fact that for the {\it single coating} there has to be boundary conditions on the inside of $\partial K$ has
already been observed by \cite{gklu}. In Definition 4.1 of \cite{gklu} a definition of finite energy solutions
is given. Furthermore, is proven in Theorem 6.1 that  in the case of  the {\it single coating} -where the permittivity and the permeability are bounded
above and below inside the cloaked object- the tangential components of  the electric and the magnetic
field of these solutions have to vanish in the inside of the
boundary of the cloaked object. Note that in this case in order to have a self-adjoint extension of the
electromagnetic propagator inside the cloaked object we are only allowed to require that either the tangential
component of $\e$ or the tangential component of $\h$ vanishes, but not both.

These boundary conditions are called {\it hidden boundary conditions} in  \cite{gklu} where also the case of the
Helmholtz equation is considered. In the case of Maxwell's equations they propose two solutions to this issue.
One of them is a lining, i.e., a physical
material on the boundary of the cloaked object that enforces a
particular boundary condition, for example, they propose  a lining by a perfect electric conductor.  Note
that this raises now the question of what is the boundary
condition between the lining and the cloaking metamaterial. In fact, we face the same problem as before, since
we can always consider that the lining is part of the cloaked objects, and then, the question of what is the
appropriate boundary condition remains.
The second proposal of \cite{gklu} is  a {\it double coating}  that corresponds to surrounding both the inner
and the outersurface of the cloaked objects with appropriately matched metamaterials.
As our permittivities
and permeabilities inside $K$  are allowed to vanish as they approach $\partial K$ the {\it double coating}
 fits in our formalism.

In Theorem 5.1 of \cite{gklu} cloaking is proven for all frequencies and active devices,  with the
{\it double coating},
 with respect to the Cauchy data of the finite energy solutions that they define in  Definition 4.1.

Remark  that there is no real contradiction between our results and the ones of \cite{gklu}. Our results
 imply that there is always a {\it hidden boundary condition}  on the inside of the boundary of the cloaked
 objects, that is imposed upon us by the fundamental principle
 of the conservation of the energy of the electromagnetic waves, that implies that time evolution has to be
 given by a unitary group generated by a self-adjoint
 extension of the electromagnetic propagator, and this amounts to  a boundary condition at the inside of the
 boundary of the cloaked objects. Note that we do not exclude here
 the possibility that in some cases  the electromagnetic propagator of the cloaked objects could be
 essentially self-adjoint, and in this situation the dynamics inside the
 cloaked objects will be uniquely defined. In this case the {\it hidden boundary condition} will be uniquely
 determined by the boundary conditions satisfied by the functions in the domain of the unique self-adjoint realization of the electromagnetic propagator in the
cloaked objects. Note, however, that we have proven that for {\it exact transformation media}
the cloaking outside is actually independent of the cloaked objects.

\section{Cloaking as a Boundary Value Problem}
\sss
It is a question of independent interest to consider cloaking as a boundary value problem for the Maxwell'
system at a fixed frequency

\begin{eqnarray}
\nabla \times \e &=&  i\lambda \b, \,\, \nabla \times \h \,=\,-i\lambda \d,\,\, \lambda \neq 0,
 \label{4.1}\\
 \nabla \cdot \b&=&0, \nabla \cdot \d\,=
 \,0.
\label{4.2}
 \end{eqnarray}

As we have already shown, cloaking is independent of the cloaked object, and this means that we only have to
consider these equation in $\Omega$. The main question now is to decide what is an appropriate class
of solutions with locally finite energy. Our analysis of the self-adjoint extensions of the electromagnetic
propagator shows that we have to take solutions that are locally in the domain of $A_\Omega$, that is to say
that they are given by (\ref{2.1})

$$
\left(\begin{array}{c}\e\\ \h\end{array}\right)= U \left(\begin{array}{c}\e_0\\ \h_0\end{array}\right)
$$

with $(\e_0,\h_0)^T$ locally in the domain of $A_0$, i.e., $(\e_0,\h_0)^T$ are in the domain of $A_0$
 when
multiplied by any function in $\C^\infty_0(\ere^3)$.
It follows from (\ref{2.2}) that the solutions with locally finite energy have to satisfy the boundary condition,

$$
\e\times \mathbf n=0, \h\times  \mathbf n=0, \,\, \hbox {in }\,\, \partial K_+,
$$

where $\partial K_+$ is the outside of the boundary of the cloaked object.  This is the only
self-adjoint boundary condition on $\partial K_+$. Note that we  define in the same way solutions with
(locally) finite energy in a bounded subset of $\Omega$. In \cite{gklu} a different definition of solutions
with (locally) finite energy is given in Definition 4.1.

\section{Cloaking an Infinite Cylinder}
We discuss now the case of an infinite cylinder. For simplicity we consider one cylinder centered at
zero and with its axis  the vertical line $L:=(0,0,x^3), x^3 \in \ere $. Then,

\beq
K:=\left\{ \x=(x^1,x^2,x^3) \in \ere^3: \sqrt{|x^1|^2+|x^2|^2}\leq a, x^3 \in \ere \right\},\,\, \Omega:= \ere^3 \setminus K.
\label{5.1}
\ene

The set $\Omega_0$ is now given by,

\beq
\Omega_0= \ere^3 \setminus L.
\label{5.2}
\ene
 Let us denote by $\X:= (x,^1,x^2)$ the vectors in the $x^1-x^2$ plane and $\hat\X:= \X/|\X|$. The
transformation (\ref{1.8}) is replaced by
\beq
\x=\x(\y)= f(\y):= \left\{ \begin{array}{c}\X= (\frac{b-a}{b}|\Y|+a) \hat{\Y},\\ x_3=y_3,\end{array} \right.
\label{5.3}
\ene
for $0 <|\Y| \leq b$ and with $ b >a$.
This transformation blows up the line $L$ onto $\partial K$ and  it sends $K_b\setminus L$ onto
$ K_b\setminus K$ where
$$
K_b:=\left\{ \y=(y^1,y^2,y^3) \in \ere^3: \sqrt{|y^1|^2+|y^2|^2}\leq b, y^3 \in \ere \right\}.
$$

For $\y \in \ere^3 \setminus K_b$ we define the transformation to be the identity, $\x= \y$.

The Hilbert spaces of finite energy electromagnetic fields, the unitary operator $U$ and  $a_0$,
$A_0, a_\Omega,  a_K, a$, are defined as in Section 2.

\begin{theorem}
The operator $a_\Omega$ is essentially self-adjoint, and its unique self-adjoint extension, $A_\Omega$,
satisfies
\beq
A_\Omega = U\, A_0\, U^\ast.
\label{5.4}
\ene
\end{theorem}
\noindent {\it Proof:}
The theorem is proven as Theorem 2.1 observing that $W^{1,2}(\ere^2)= W^{1,2}_0(\ere^2\setminus 0)$
since $\{0\}$ has zero Sobolev one capacity in $\ere^2$ \cite{af,kkm,km}.

\bull

We now consider the wave and the scattering operators. For simplicity we assume below that
$\varepsilon_0^{\lambda\nu}= \tilde{\varepsilon}\, \delta^{\lambda\nu}, \mu_0^{\lambda\nu}= \tilde{\mu}\,
\delta^{\lambda\nu}$.
The wave operators are defined as in (\ref{1.56}) but now the operator $J$ is defined as follows,

$$
\left(J \left(\begin{array}{c}\e_0\\ \h_0\end{array}\right)\right)(\x):=  \chi_\Omega(\x) \phi(\underline{\x})
\left(\begin{array}{c}\e_0\\ \h_0\end{array}\right)(\x),
$$
where $ \phi$ is continuous and it  satisfies $\phi(\underline{\x})= (|\underline{\x}|-a), a \leq
|\underline{\x}| \leq a+\delta$ and $\phi(\underline{\x})=1$ for $|\underline{\x}| \geq a+2 \delta$, for
some $ \delta >0$.

\begin{lemma}
\beq
W_{\pm}= U P_{0\perp}.
\label{5.5}
\ene
\end{lemma}
\noindent {\it Proof:} The lemma is proven as in the proof of Lemma 2.2, but in (\ref{2.4},
 \ref{2.5}, \ref{2.6}, \ref{1.59}) we have to replace $\chi_{\ds B_R}$, by
$\chi_{\ds C_R}$ where, $C_R:=\{ \y \in \ere^3: |\Y|\leq R\}$ for $R$ large enough. Now we can not prove
(\ref{2.7}, \ref{2.8}, \ref{1.60}) by compactness arguments because $K$ is unbounded. Instead we use
propagation estimates for $A_0$. The following results are well known. See for example \cite{ch,we1,we2,wi}
where the general anisotropic case is considered. For any $\varphi \in \H_{0\perp}$,

\beq
e^{-it A_0}\varphi= \frac{1}{\ds (2\pi)^{3/2}}\int_{\ere^3}  e^{ik\cdot\y}  \left( e^{-i \omega_+(k)t}
P_+(k) \hat{\varphi}(k)+  e^{-i \omega_-(k)t}
P_-(k) \hat{\varphi}(k)\right)\, d^3 k
\label{5.6}
\ene
where $\hat{\varphi}$ is the Fourier transform of $\varphi$, $\omega_\pm(k)=\pm |k| c$ with $c:=
 (\tilde{\varepsilon}
\tilde{\mu})^{-1/2}$, and $P_\pm(k)$ are projectors on $\ere^3$ that are infinitely differentiable for $ k \in
\ere^3\setminus 0$. Suppose that $ \hat{\varphi} \in \mathbf C^\infty_0(\ere^3 \setminus L)$ and let $O$ be a
bounded open set such that
$\overline{O} \subset \ere^3 \setminus L$ and $\hbox {support}\, \hat{\varphi} \subset O$.
Denote
$$
\hat{O}:=\left\{ \frac{k}{|k|}: k \in O\right\}.
$$
Then by the (non) stationary phase Theorem (see the Corollary to Theorem XI.14 of \cite{rsIII} ), for any $  n=1,2,\cdots$ there is a constant $C_n$ such that
\beq
\left|\left(e^{-it A_0}\varphi\right)(\y)\right| \leq C_n \left(1+|\y|+|t|\right)^{-n}, \, \, \pm
\frac{\y}{ct} \notin \hat{O}.
\label{5.7}
\ene

We write,

\beq
\chi_{ C_R} e^{-it A_0}\varphi = \phi_1 +\phi_2
\label{5.8}
\ene
with
\beq
\phi_1:= \chi_{(\pm \y/(c t) \notin \hat{O})} \chi_{ C_R} e^{-it A_0}\varphi
\label{5.9}
\ene
and

\beq
\phi_2:= \chi_{(\pm \y/(ct) \in \hat{O})} \chi_{ C_R} e^{-it A_0}\varphi.
\label{5.10}
\ene

by (\ref{5.7})

\beq
\hbox{s-}\lim_{t\rightarrow \pm \infty} \phi_1=0.
\label{5.11}
\ene

Note, furthermore, that there is an $\epsilon >0$ such that $|\underline{k}| \geq \epsilon$ for any
$k \in \hat{O}$. Then, for any $ \pm \frac{\y}{c t}\in \hat{O}, |\underline{\y}| \geq c |t|\epsilon$.
It follows that there is a $T$ such that

\beq
\phi_2 =0, \, \hbox{for} |t| \geq T.
\label{5.12}
\ene

By (\ref{5.8}, \ref{5.11}, \ref{5.12})

\beq
\hbox{s-}\lim_{t\rightarrow \infty} \chi_{ C_R} e^{-it A_0}\varphi=0,
\label{5.13}
\ene

and (\ref{2.8}, \ref{1.60}) follow. Note that $P_{0\perp}$ is not needed because $ \varphi \in \H_{0\perp}$.
 By continuity this is true for all $\varphi \in \H_{0\perp}$. Then, (\ref{5.5}) follows from (\ref{1.58},
 \ref{1.60}).

\bull

The scattering operator is defined as in (\ref{2.9}).
\begin{corollary}
\beq
S=P_{0\perp}.
\label{5.14}
\ene
\end{corollary}
\noindent {\it Proof:} This is immediate from (\ref{5.5}) because $U^\ast\, U=I$.

\bull

Let us denote by $S_\perp$ the restriction of $S$ to $\H_{0\perp}$. $S_{\perp}$ is the physically relevant scattering operator that acts in the
Hilbert space $\H_{0\perp}$ of finite energy fields that satisfy equations (\ref{1.2}). We designate by $I_\perp$ the identity operator on
$\H_{0\perp}$. We have that,

\begin{corollary}
\beq
S_\perp= I_\perp.
\label{5.15}
\ene
\end{corollary}
\noindent {\it Proof:} This follows from Corollary 5.3.

\bull

Again, the fact that $S_\perp$ is the identity operator on $\H_{0\perp}$  means that there is cloaking for all
 frequencies.

In Theorem 7.1 of \cite{gklu} cloaking is proven for all frequencies with respect to the Cauchy data of the
finite energy solutions that they define in Definition 4.1 and furthermore, in Theorem 8.2, they prove cloaking for all frequencies
with the SHS boundary condition  with respect to the Cauchy data of the finite energy solutions that they define
in Definition 8.1.

  Theorem 3.1 remains true in the case of the cylinder. The proof is the same.
Furthermore, all the remarks about finite energy solutions, and cloaking and  that we made in Sections 2, 3,
are true in the case of a cylinder. We do not repeat them here.
Moreover, equations (\ref{2.1}) hold. However, since now the transformation (\ref{5.3}) only acts on the plane orthogonal
to the axis of the cylinder equations (\ref{2.2}) has to be replaced by

\beq
\underline{\e}\times \hat{\underline{\x}}=0, \, \underline{\h }\times \hat{\underline{\x}}=0,\, \hbox{in}\,
\partial K_+,
\label{5.16}
\ene
where $\underline{\e}:=(E_1,E_2), \underline{\h}:=(H_1,H_2)$.

 As in Section 4 we define solutions to (\ref{4.1}, \ref{4.2}) with locally finite energy as solutions
  that are locally in the domain of $A_\Omega$, that is to say
that they are given by (\ref{2.1})

$$
\left(\begin{array}{c}\e\\ \h\end{array}\right)= U \left(\begin{array}{c}\e_0\\ \h_0\end{array}\right)
$$

with $(\e_0,\h_0)^T$ locally in the domain of $A_0$, i.e., $(\e_0,\h_0)^T$ are in the domain of $A_0$
 when
multiplied by any function in $\C^\infty_0(\ere^3)$.
It follows from (\ref{5.16}) that the solutions with locally finite energy have to satisfy the boundary
condition,

\beq
\underline{\e}\times \hat{\underline{\x}}=0, \, \underline{\h }\times \hat{\underline{\x}}=0,\, \hbox{in}\,
\partial K_+.
\label{5.17}
\ene

Note that (\ref{5.17}) is the SHS boundary condition considered  in \cite{gklu}.
We have proven here that (\ref{5.17}) is the only self-adjoint boundary condition on $\partial K_+$
allowed by energy conservation.

\noindent {\bf Acknowledgement}

\noindent This work was partially done while I was visiting the Institut f\"ur Theoretische Physik,
Eidgen\"ossische Techniche H\"ochschule Zurich. I thank professors Gian Michele Graf and J\"urg
Fr\"ohlich for their kind hospitality.

\end{document}